\documentclass{PoS}
\usepackage{colordvi}
\usepackage{epsfig}

\title{Renormalization of Flavor Singlet and Nonsinglet Fermion
Bilinear Operators}

\ShortTitle{Renormalization of Flavor Singlet and Nonsinglet
Fermion Bilinear Operators}

\author{Martha Constantinou$^a$, Marios Hadjiantonis$^{a,b}$, \speaker{Haralambos Panagopoulos}\,$^a$\\
\llap{$^a$} Department of Physics, University of Cyprus, Nicosia
CY-1678, Cyprus\\
\llap{$^b$} Present address: Department of Physics, University of Michigan, Ann Arbor, MI 48109, USA\\
        E-mail: \email{constantinou.martha@ucy.ac.cy},
\email{mchatz12@ucy.ac.cy}, \email{haris@ucy.ac.cy}}

\abstract{We compute the difference in the renormalization of flavor
singlet and nonsinglet fermion bilinear operators, to two loops in
perturbation theory. Our results are applicable to a rather wide class
of lattice actions with Symanzik improved gluons, stout links and clover
fermions, including the Twisted Mass and SLiNC actions.

A more detailed presentation of our results, along with 
relevant references, will appear in our
forthcoming publication \cite{CHP}.}

\FullConference{The 32nd International Symposium on Lattice Field Theory,\\
		23-28 June, 2014\\
		Columbia University New York, NY}

\newcommand{\be}{\begin{equation}}
\newcommand{\ee}{\end{equation}}
\newcommand{\bea}{\begin{eqnarray}} 
\newcommand{\eea}{\end{eqnarray}}
\def\slashed{{/}\mskip-10.0mu}

\begin{document}

\section{Introduction}
In this work we study the renormalization of fermion bilinears 
${\cal O}_\Gamma = \bar{\psi}\Gamma\psi$ on the lattice, where  
$\Gamma = \hat{1},\,\gamma_5,\,\gamma_{\mu},
\,\gamma_5\,\gamma_{\mu},\, \gamma_5\,\sigma_{\mu\,\nu}$.  
We consider flavor
singlet ($\sum_f\bar\psi_f\Gamma\psi_f$, f: flavor index) as well as
nonsinglet operators, to two
loops in perturbation theory.
Our calculations were performed making use of a large family of
lattice actions, including Symanzik improved gluons, Wilson
fermions with stout links, and clover fermions; twisted
mass actions (with Iwasaki or tree level Symanzik gluons) and
the SLiNC action are members of this family. 

The most demanding parts of this study are the computation of the
2-point Green's functions of ${\cal O}_\Gamma$\,, up to two loops and,
as a prerequisite, the two-loop fermion self-energy. From
these Green's 
functions we extract the renormalization functions for the
fermion field and for ${\cal O}_\Gamma\,$: 
$Z_\psi^{L,Y}$, $Z_\Gamma^{L,Y}$ ($L$: lattice regularization, $Y$ 
$(= RI^{\prime},\ \overline{MS})$: renormalization schemes). As a
check on our results, we have computed them in an arbitrary covariant
gauge. Our expressions for $Z_\psi^{L,Y}$, $Z_\Gamma^{L,Y}$ can be
generalized, in a straightforward manner, to fermionic fields in an
arbitrary representation.

Flavor singlet operators are relevant for a number of hadronic
properties including, e.g., topological features or the spin structure
of hadrons. Matrix elements of such operators are notoriously
difficult to study via numerical simulations, due to the presence of
(fermion line) disconnected diagrams, which in principle require
evaluation of the full fermion propagator. In recent years
there has been some progress in the numerical study of flavor singlet
operators; furthermore, for some of them, a nonperturbative estimate
of their renormalization has been obtained using the Feynman-Hellmann
relation~\cite{FH}. Perturbation theory can give an important cross
check for these estimates, and provide results for
operators which are more difficult to renormalize nonperturbatively.

Given that for the renormalization of flavor nonsinglet operators one
can obtain quite accurate nonperturbative estimates, we will focus on the
perturbative evaluation of the \emph{difference} between the
flavor singlet and nonsinglet renormalization; this difference first
shows up at two loops.

Perturbative computations beyond one loop for Green's functions
with nonzero external momenta are technically quite involved, and
their complication is greatly increased when improved gluon and
fermion actions are employed. For fermion bilinear operators, the only
two-loop computations in standard perturbation theory thus far have been
performed by our group~\cite{SP}, employing Wilson gluons and
Wilson/clover fermions. Similar investigations have been carried out
in the context of stochastic perturbation theory~\cite{DiRenzo}.

Further composite fermion operators of interest, to which one can apply our
perturbative techniques, are higher dimension bilinears such as:
$\bar{\psi}\Gamma\,D^\mu\psi$ (appearing in hadron structure
functions) and four-fermion operators such as: 
$(\bar{s}\,\Gamma_1\,d)\,(\bar{s}\,\Gamma_2\,d)$ (appearing in $\Delta
S = 2$ transitions, etc.); in these cases, complications such as
operator mixing greatly hinder nonperturbative methods of
renormalization, making a perturbative approach all that more essential.

\section{Definitions and Calculational Setup}

In standard notation, the action consists of a 
gluon part, $S_G$\,, and a fermion part, $ S_W + S_{SW}$\,:
\bea
\hskip -1cm S_W&=&-\frac{a_{_{\rm L}}^3}{2}\sum_{n,\,\mu}\Big[\overline{\Psi}(\vec{n})\Big((r-\gamma_{\,\mu})\Red{{\tilde U}_{\vec{n},\,\vec{n}+\hat{\mu}}}\Psi(\vec{n}+\hat{\mu})
+(r+\gamma_{\,\mu})\Red{{\tilde
    U}^{\dagger}_{\vec{n}-\hat{\mu},\,\vec{n}}}\Psi(\vec{n}-\hat{\mu})
-2r\Psi(\vec{n})\Big)\Big] \\
S_{SW}&=&-a_{_{\rm L}}^5\sum_{n}
\MidnightBlue{c_{SW}}\,\overline{\Psi}(\vec{n})\,\frac{1}{4}\,\sigma_{\mu\nu}\,\hat{G}_{\mu\nu}(\vec{n})\,\Psi(\vec{n}),\qquad
\hat{G}_{\mu\nu}(\vec{n})=[Q_{\mu\nu}(\vec{n})-Q_{\nu\mu}(\vec{n}) ]/(8a_{_{\rm L}}^2),\eea
\bea \hskip -1cm Q_{\mu\nu}&=&\,\phantom{+}
U_{\vec{n},\,\vec{n}+\hat{\mu}}\,U_{\vec{n}+\hat{\mu},\,\vec{n}+\hat{\mu}+\hat{\nu}}\,U_{\vec{n}+\hat{\mu}+\hat{\nu},\,\vec{n}+\hat{\nu}}\,U_{\vec{n}+\hat{\nu},\,\vec{n}}+\,U_{\vec{n},\,\vec{n}+\hat{\nu}}\,U_{\vec{n}+\hat{\nu},\,\vec{n}+\hat{\nu}-\hat{\mu}}\,U_{\vec{n}+\hat{\nu}-\hat{\mu},\,\vec{n}-\hat{\mu}}\,U_{\vec{n}-\hat{\mu},\,\vec{n}}
\nonumber \\
&&+\,U_{\vec{n},\,\vec{n}-\hat{\mu}}\,U_{\vec{n}-\hat{\mu},\,\vec{n}-\hat{\mu}-\hat{\nu}}\,U_{\vec{n}-\hat{\mu}-\hat{\nu},\,\vec{n}-\hat{\nu}}\,U_{\vec{n}-\hat{\nu},\,\vec{n}}
+\,U_{\vec{n},\,\vec{n}-\hat{\nu}}\,U_{\vec{n}-\hat{\nu},\,\vec{n}-\hat{\nu}+\hat{\mu}}\,U_{\vec{n}-\hat{\nu}+\hat{\mu},\,\vec{n}+\hat{\mu}}\,U_{\vec{n}+\hat{\mu},\,\vec{n}}\eea
The fermion action may also
contain standard and twisted mass terms, but they
only contribute beyond two loops to the difference between flavor singlet
and nonsinglet renormalizations; this is true in mass-independent
schemes, such as $RI^{\prime}$ and $\overline{MS}$, in which renormalized
masses vanish. The quantities $\Red{{\tilde
    U}_{\vec{n},\,\vec{n}+\hat{\mu}}}$, appearing above, are stout
links, defined as: ${\displaystyle \Red{\tilde{U}_{\vec{n}, \vec{n} +
      \hat{\mu}}} = e^{\displaystyle i Q_{\hat{\mu}} (\vec{n})}
  U_{\vec{n}, \vec{n} + \hat{\mu}}}$\,, where:
\be Q_{\hat{\mu}} (\vec{n}) = \frac{\MidnightBlue{\omega}}{2 i}  \bigl[ V_{\hat{\mu}} (\vec{n}) U_{\vec{n}, \vec{n} + \hat{\mu}}^\dagger - U_{\vec{n}, \vec{n} + \hat{\mu}} V_{\hat{\mu}}^\dagger(\vec{n}) - \frac{1}{N_c} {\rm Tr} \bigl(V_{\hat{\mu}}(\vec{n}) U_{\vec{n}, \vec{n} + \hat{\mu}}^\dagger - U_{\vec{n}, \vec{n} + \hat{\mu}} V_{\hat{\mu}}^\dagger(\vec{n})\bigr) \Bigr]\ee
and $V_{\hat{\mu}}(\vec{n})$ is the sum of 6
staples joining sites $\vec n$ and $\vec n + \hat\mu$. Both the stout
coefficient 
$\MidnightBlue{\omega}$ and the clover coefficient $\MidnightBlue{c_{SW}}$ will be treated as free parameters, for wider
applicability of the results.

We employ a Symanzik improved gluon action, of the form:
\be \hskip 0.02cm S_G = \frac{2}{g^2}  {\rm Re}\, {\rm Tr} \Big[ \MidnightBlue{c_0} \sum_\mathrm{Plaq} (1 - U_\mathrm{Plaq}) + \MidnightBlue{c_1} \sum_\mathrm{Rect} (1 - U_\mathrm{Rect}) + \MidnightBlue{c_2} \sum_\mathrm{Chair} (1 - U_\mathrm{Chair}) + \MidnightBlue{c_3} \sum_\mathrm{Paral} (1 - U_\mathrm{Paral}) \Big]\ee
where:
$U_\mathrm{Plaq}\ (U_\mathrm{Rect},\ U_\mathrm{Chair},\ U_\mathrm{Paral})$
is the product of links around a $1{\times}1$ plaquette (the three
independent 6-link Wilson loops: ``$2{\times}1$ rectangle'',
``chair'', ``parallelogram''), and the Symanzik coefficients $c_i$ satisfy: 
$\MidnightBlue{c_0} + 8 \MidnightBlue{c_1} + 16 \MidnightBlue{c_2} + 8 \MidnightBlue{c_3} = 1$.
The algebraic part of our computation was carried out for generic
values of $c_i$\,; for the numerical integration over loop momenta we
selected a number of commonly used sets of values, some of which are
shown in Table~\ref{tabSymanzik}.
\begin{table}[ht]
\centering
\begin{tabular}{ccccc}
\hline\hline
Action & $c_0$& $c_1$ & $c_2$ & $c_3$ \\
\hline
Wilson & $1$ & $0$ & $0$ & $0$ \\
Tree-Level Symanzik & $5/3$ & $-1/12$ & $0$ & $0$ \\
Iwasaki & $3.648$ & $-0.331$ & $0$ & $0$ \\
DBW2 & $12.2688$ & $-1.4086$ & $0$ & $0$ \\
\hline\hline
\end{tabular}
\caption{Selected sets of values for Symanzik coefficients}
\label{tabSymanzik}
\end{table}

Denoting all bare quantities in the Lagrangian with the subscript ``$\circ$'',
the corresponding renormalized quantities and
renormalization functions read:
\be A_{\mu\,\circ}^a = \sqrt{Z_A}\,A^a_{\mu}, \hspace{0.5cm}
c^a_\circ=\sqrt{Z_c}\,c^a, \hspace{0.5cm}
\psi_\circ=\sqrt{Z_{\psi}}\,\psi, \hspace{0.5cm}
g_\circ = \mu^{\epsilon}\,Z_g\,g, \hspace{0.5cm} \alpha_\circ =
Z_\alpha^{-1}\,Z_A\,\alpha \ee
(respectively: gauge field, ghost field, fermion field, coupling constant, gauge parameter)
where $\mu$ is the mass scale introduced to ensure that $g_\circ$ has
the correct dimensionality in $d=4-2\epsilon$ dimensions. $Z_A$,
$Z_c$, $Z_g$, $Z_\alpha$ are only needed to one loop, and their values
are known.

Aside from $c_{SW}$ and $\omega$, we treat as free parameters: $g_0$,
$\alpha$, $N_f\,(N_c)$\,: number of flavors (colors). 

The 2-point amputated Green's functions of the
operators ${\cal O}_\Gamma$\, can be written in the form:
\bea
\Sigma_S(q a_{_{\rm L}})&=&\hat{1}\,\Sigma^{(1)}_S (q a_{_{\rm L}})\label{GreenS}\\
\Sigma_P(q a_{_{\rm L}})&=&\gamma_5\,\Sigma^{(1)}_P (q a_{_{\rm L}})\\
\Sigma_V(q a_{_{\rm L}})&=&\gamma_\mu\,\Sigma^{(1)}_V (q a_{_{\rm L}}) +
q^\mu(\slashed q/q^2)\Sigma^{(2)}_V (q a_{_{\rm L}}) \\
\Sigma_{AV}(q a_{_{\rm L}})&=&\gamma_5\gamma_\mu\,\Sigma^{(1)}_{AV} (q a_{_{\rm L}}) + 
\gamma_5q^\mu(\slashed q/q^2)\Sigma^{(2)}_{AV} (q a_{_{\rm L}}) \label{GreenAV}\\
\Sigma_T(q a_{_{\rm L}})&=&\gamma_5\,\sigma_{\mu\,\nu}\Sigma^{(1)}_T(q a_{_{\rm L}}) + 
\gamma_5(\slashed q/q^2)\,(\gamma_\mu q_\nu - \gamma_\nu
  q_\mu)\Sigma^{(2)}_T(q a_{_{\rm L}})\label{GreenT}
\eea
$\big(S,\, P,\, V,\, AV,\, T$ correspond to: $\Gamma =
\hat{1},\, \gamma_5\,,\, \gamma_\mu\,,\, \gamma_5\gamma_\mu\,,
\, \gamma_5\,\sigma_{\mu\,\nu}$\,. 
$\Sigma^{(1)}_\Gamma= 1 + {\cal
  O}(g_\circ^2)$, $\Sigma^{(2)}_\Gamma= {\cal O}(g_\circ^2)\big)$.

\bigskip
The $RI'$ renormalization scheme is defined by
imposing renormalization conditions on matrix elements at a scale
$\bar{\mu}$, where (just as in $\overline{MS}$): 
$\bar{\mu}=\mu\,(4\pi/e^{\gamma_{\rm E}})^{1/2}$.
For the renormalized operators: 
${\cal O}^{RI^{\prime}}_{\Gamma}=Z^{L,RI^{\prime}}_{\Gamma}\!(a_{_{\rm L}}\bar\mu)
\,{\cal O}_{\Gamma\,\circ}\ ({\cal
  O}_{\Gamma\,\circ}=\bar{\psi}\,\Gamma\,\psi)$, one defines
$Z_\Gamma^{L,RI^{\prime}}$ via:
\be\lim_{a_{_{\rm L}}\rightarrow
  0}\left[Z_{\psi}^{L,RI^{\prime}}\,\Red{Z_\Gamma^{L,RI^{\prime}}}\,\Sigma^{\Red{(1)}}_\Gamma
  (q a_{_{\rm L}})\right]_{q^2=\bar{\mu}^2} = 1\ee
This renormalization prescription does not involve
$\Sigma^{\Red{(2)}}_\Gamma$\,; nevertheless, renormalizability of the
theory implies that $Z_\Gamma^{L,RI^{\prime}}$ will render the entire
Green's function finite. An alternative prescription, more
appropriate for nonpertubative renormalization is:
\be\lim_{a_{_{\rm L}}\rightarrow 0}
\left[Z_{\psi}^{L,RI^{\prime}}\,\Red{Z_\Gamma^{L,RI^{\prime}}}\, {\rm
      tr}(\Gamma\Sigma_\Gamma (q a_{_{\rm L}})) / {\rm
      tr}(\Gamma\Gamma)\right]_{q^2=\bar{\mu}^2} = 1\ee
The two prescriptions differ between themselves (for V, AV, T) by
a finite amount which can be deduced from lower loop
calculations combined with continuum results. 

Conversion of renormalization functions from $RI^{\prime}$ to
the $\overline{MS}$ scheme is facilitated by the fact
that renormalized Green's 
functions are regularization independent; thus
the finite conversion factors
($DR\equiv$ Dimensional Regularization): 
\be C_\Gamma(g,\alpha)\equiv \frac{Z_\Gamma^{L,RI^{\prime}}}{Z_\Gamma^{L,\overline{MS}}}
=\frac{Z_\Gamma^{DR,RI^{\prime}}}{Z_\Gamma^{DR,\overline{MS}}}\ee
can be evaluated in DR, leading to:
$Z_\Gamma^{L,\overline{MS}}=Z_\Gamma^{L,RI^{\prime}} /
C_\Gamma(g,\alpha)$. 
For the pseudoscalar and axial vector operators, in order to satisfy Ward identities, additional finite factors $Z_5^P(g)$ and $Z_5^{AV}(g)$, calculable in DR, are required: 
\be Z_P^{L,\overline{MS}} = Z_P^{L,RI^{\prime}} / \left( C_S\,Z_5^P \right),\quad
Z_{AV}^{L,\overline{MS}} = Z_{AV}^{L,RI^{\prime}} / \left( C_V\,Z_5^{AV} \right)\ee
These factors are gauge invariant; we also note that the value of $Z_5^{AV}$ for the flavor singlet operator differs from that of the nonsinglet one.

\section{Computation and Results}
In the previous Section, the calculational setup was presented in
rather general terms. Here we focus on the two-loop difference between
flavor singlet and nonsinglet operator renormalization. Given that
this difference first arises at two loops, we only need the tree level
values of: $Z_\psi$, $Z_A$, $Z_c$, $Z_\alpha$, $Z_g$ and of the
conversion factors $C_\Gamma$, $Z_5^P$ and $Z_5^{AV}$. Since the tree
level value of $C_\Gamma$ equals 1, the difference up to two loops
will not depend on the renormalization scheme.

\begin{figure}[h]
\begin{center}
\includegraphics[width=.6\textwidth]{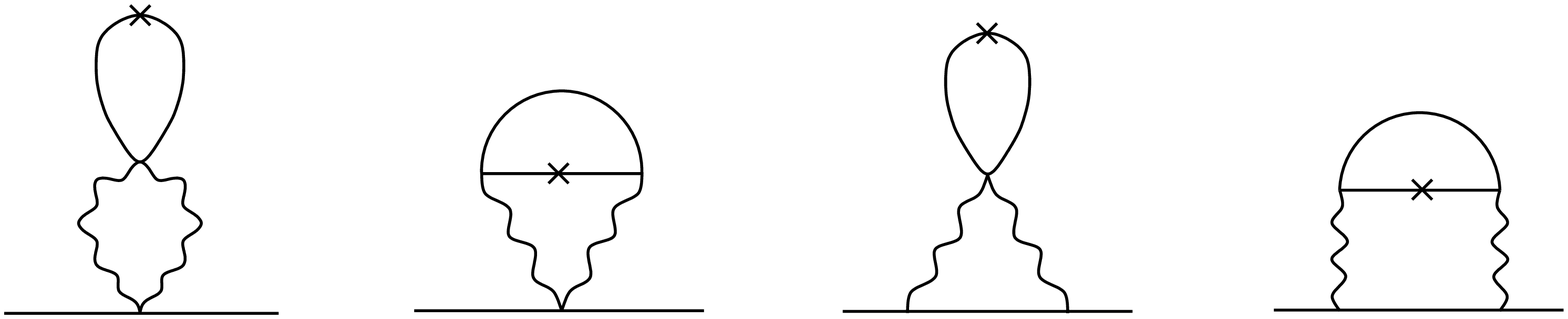}
\caption{Diagrams contributing to the difference between flavor
  singlet and nonsinglet values of $Z_\Gamma$\,. Solid (wavy) lines
  represent fermions (gluons). A cross denotes insertion of the
  operator $\sum\bar\psi\Gamma\psi$.}
\label{Feynman}
\end{center}
\end{figure}

There are 4 two-loop Feynman diagrams contributing to the above
difference in the evaluation of the Green's
functions~(\ref{GreenS}-\ref{GreenT}), shown in
Fig.~\ref{Feynman}. They all contain an operator insertion inside a closed
fermion loop, and therefore vanish in the flavor nonsinglet case.

The above diagrams, evaluated individually, may be IR divergent, due to
the presence of two gluon propagators with the same
momentum. Comparing to our previous evaluation of these diagrams with
Wilson gluons and clover fermions, we will find no new superficial
divergences, i.e. no new ${\rm ln}^2(\bar\mu a_{_{\rm L}})$ terms.
However, the presence of stout links and Symanzik gluons leads to
considerably longer expressions for the vertices. Also, the gluon
propagator must now be inverted numerically for each value of the
loop momentum 4-vector (an inversion in closed form exists, but it is
not efficient).

The contribution of these diagrams to $Z_P$, $Z_V$,
$Z_T$ vanishes identically. Only $Z_S$ and $Z_{AV}$
are affected.
For the Scalar operator, our result can be written in the following form:
\bea Z_S^{\rm singlet}\! (\bar\mu a_{_{\rm L}}) &-& Z_S^{\rm
  non-singlet}\! (\bar\mu a_{_{\rm L}}) = - \frac{g_0^4}{\left( 4 \pi \right)^4} c_F N_f \bigl[
  \bigl(s_{00} + s_{01}\ \Red{\mathbf c_{SW}} + s_{02}\ \Red{\mathbf
    c_{SW}^2}  + s_{03}\ \Red{\mathbf c_{SW}^3} + s_{04}\ \Red{\mathbf
    c_{SW}^4} \bigr) \nonumber\\
&&\hspace*{.2cm} + \bigl(s_{10}\ + s_{11}\ \Red{\mathbf c_{SW}} +
s_{12}\ \Red{\mathbf c_{SW}^2} + s_{13}\ \Red{\mathbf c_{SW}^3} \bigr)
\,\Blue{\mathbf \omega} + \bigl( s_{20}\ + s_{21}\ \Red{\mathbf c_{SW}} + s_{22}\ \Red{\mathbf c_{SW}^2} \bigr)  \,\Blue{\mathbf \omega^2} \nonumber\\
&&\hspace*{.2cm} + \bigl( s_{30}\ + s_{31}\ \Red{\mathbf c_{SW}} \bigr) \,\Blue{\mathbf \omega^3}  +s_{40} \, \Blue{\mathbf \omega^4} \ \bigr] + \mathcal{O}\! \left( g_0^6 \right)\label{expressionS}\eea
($c_F\equiv (N_c^2-1)/(2\,N_c)$). The numerical constants $s_{i,j}$
have been computed for various sets of values of the Symanzik
coefficients; their values are listed in Table~2. The
computation was performed in a general covariant gauge, confirming
that the result is gauge independent, as it should be in $\overline{MS}$.
We note from Eq.~(\ref{expressionS}) that even single logarithms are
absent; thus the result is scale independent.

\begin{table}[ht]
\begin{center}
\begin{tabular}{|c|ccc||c|ccc|}
\hline
 & Wilson & TL Symanzik & Iwasaki & & Wilson & TL Symanzik & Iwasaki  \\
\hline
$s_{00}$ & 107.76(2) & 76.29(1) & 42.973(7)     & $a_{00}$ & 2.051(2) & 3.098(3) & 5.226(4) \\
$s_{01}$ & -82.27(1) & -69.01(1) & -49.356(8)   & $a_{01}$ & -15.033(3) & -12.851(3) & -9.426(3) \\
$s_{02}$ & 29.730(2) & 26.178(1) & 20.312(3)    & $a_{02}$ & -5.013(2) & -3.361(1) & -1.3526(7) \\
$s_{03}$ & -3.4399(7) & -2.9533(5) & -2.2166(3) & $a_{03}$ & 2.1103(3) & 1.7260(1) & 1.1251(2) \\
$s_{04}$ & -2.2750(4) & -1.6403(3) & -0.8547(2) & $a_{04}$ & 0.0434(2) & 0.01636(1) & -0.01074(5) \\
$s_{10}$ & -1854.4(2) & -1107.0(1) & -444.69(4) & $a_{10}$ & 43.75(1) & 36.66(1) & 25.827(9) \\
$s_{11}$ & 506.26(5) & 364.01(3) & 192.35(1)    & $a_{11}$ & 76.993(3) & 57.190(3) & 31.768(2) \\
$s_{12}$ & -95.42(2) & -70.94(1) & -40.162(6)   & $a_{12}$ & 44.260(4) & 29.363(2) & 12.962(1) \\
$s_{13}$ & 7.494(1) & 5.356(1) & 2.8030(4)      & $a_{13}$ & -4.4660(6) & -3.3740(5) & -1.8710(2) \\
$s_{20}$ & 18317(2) & 10081(1) & 3511.3(4)      & $a_{20}$ & -126.45(1) & -92.853(7) & -50.378(1) \\
$s_{21}$ & -2061.8(2) & -1350.7(1) & -595.79(7) & $a_{21}$ & -259.59(3) & -175.65(2) & -81.45(1) \\
$s_{22}$ & 202.75(7) & 133.19(4) & 59.25(2)     & $a_{22}$ & -107.48(1) & -67.737(8) & -27.500(3) \\
$s_{30}$ & -96390(10) & -50300(5) & -16185(2)   & $a_{30}$ & 295.76(3) & 198.78(2) & 90.96(1) \\
$s_{31}$ & 3784.8(4) & 2336.0(3) & 925.6(1)     & $a_{31}$ & 400.05(5) & 253.87(3) & 104.74(1) \\
$s_{40}$ & 213470(20) & 106940(10) & 32572(3)   & $a_{40}$ & -348.41(4) & -220.12(3) & -90.11(1) \\
\hline
\end{tabular}
\caption{Numerical coefficients for the Scalar and Axial Vector operators. The errors
  quoted stem from numerical integration over loop momenta.} 
\end{center}
\label{resultsSandAV}
\end{table}

For the Axial Vector operator we find:
\bea Z_{AV}^{\rm singlet}\! (\bar\mu a_{_{\rm L}}) &-& Z_{AV}^{\rm
  non-singlet}\! (\bar\mu a_{_{\rm L}}) = - \frac{g_0^4}{\left( 4 \pi \right)^4} c_F N_f \bigl[ \bigl(
  a_{00} + a_{01}\ \Red{\mathbf c_{SW}} + a_{02}\ \Red{\mathbf
    c_{SW}^2} + a_{03}\ \Red{\mathbf c_{SW}^3} + a_{04}\ \Red{\mathbf
    c_{SW}^4} \bigr) \nonumber \\
&&\hspace*{.2cm} + \bigl( a_{10}\ + a_{11}\ \Red{\mathbf c_{SW}} + a_{12}\ \Red{\mathbf c_{SW}^2} + a_{13}\ \Red{\mathbf c_{SW}^3} \bigr) \,\Blue{\mathbf \omega} 
+ \bigl( a_{20}\ + a_{21}\ \Red{\mathbf c_{SW}} + a_{22}\ \Red{\mathbf
  c_{SW}^2} \bigr) \,\Blue{\mathbf \omega^2} \nonumber \\
&&\hspace*{.2cm} + \bigl( a_{30}\ + a_{31}\ \Red{\mathbf c_{SW}}
\bigr) \,\Blue{\mathbf \omega^3} + a_{40} \,\Blue{\mathbf \omega^4} +
\Magenta{6 \ln (\bar\mu^2 a_{_{\rm L}}^2)}\ \bigr] + \mathcal{O}\!
\left( g_0^6 \right) \label{expressionAV}\eea

By analogy with the scalar case, the computation was performed in a general
gauge and the numerical constants are tabulated in
Table~2. We note that the result for the Axial Vector
operator has a scale dependence; this is related to the axial anomaly. 
Finally, the presence of a term of the form 
$\gamma_5\, q^\mu\slashed
q/q^2$ in the Green's function~(\ref{GreenAV}) implies that, in the
alternative $RI'$ scheme mentioned in Section 2, one must
add $(g_0^4/(4\pi)^4)\, c_F\, N_f$ to the above result.

\begin{figure}[h]
\begin{center}
\epsfig{figure=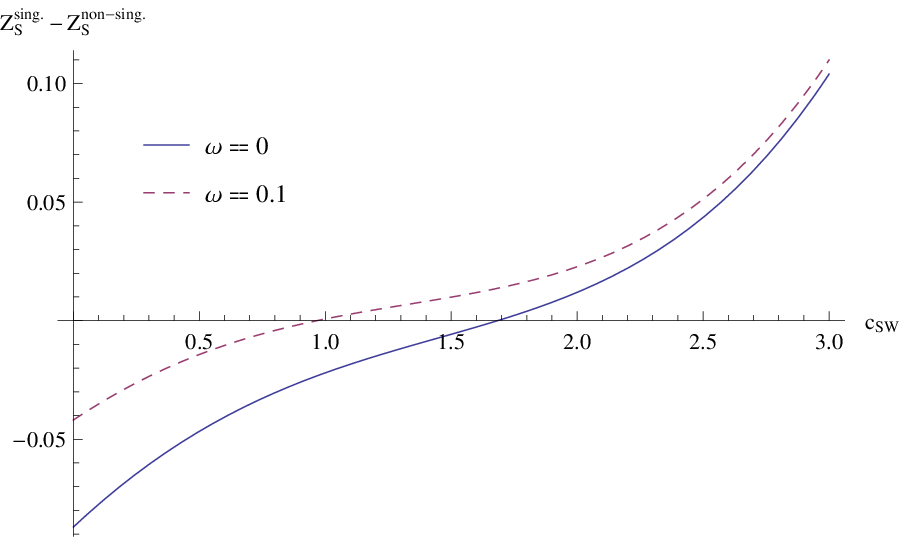,scale=0.6}
\hspace{.05\textwidth}
\epsfig{figure=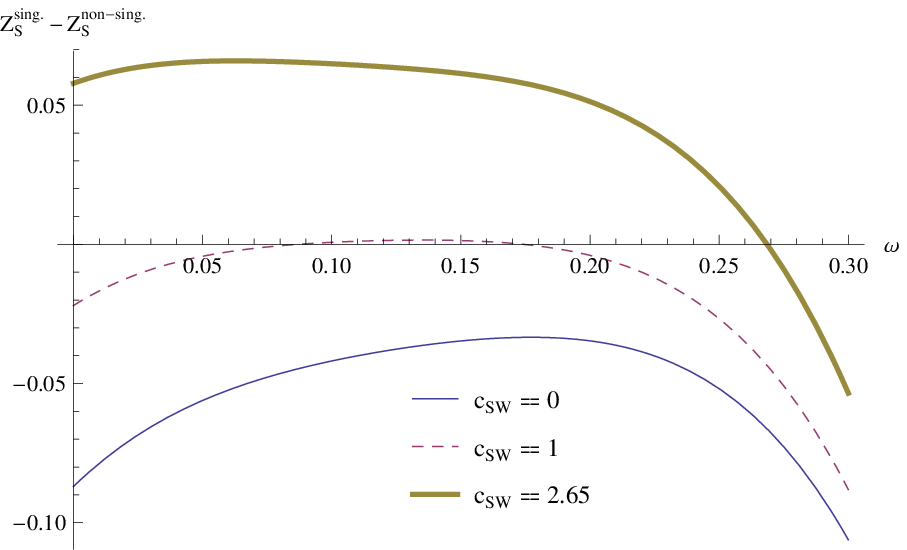,scale=0.6}\\[2mm]
\epsfig{figure=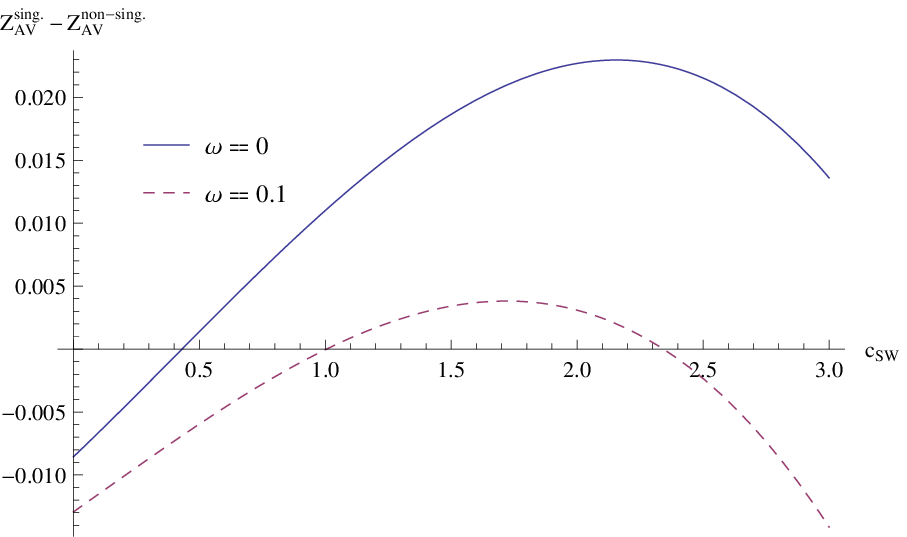,scale=0.6}
\hspace{.05\textwidth}
\epsfig{figure=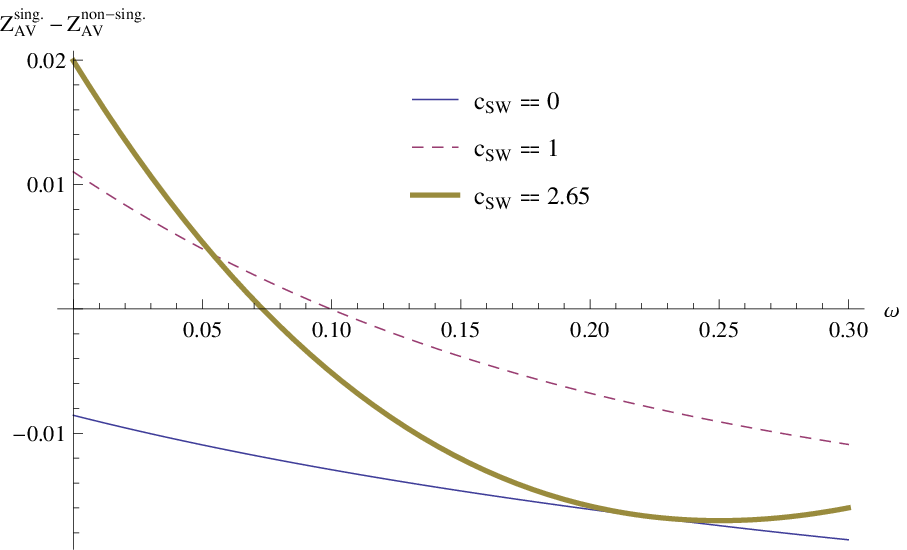,scale=0.6}

\caption{Plots of $Z_\Gamma^{\rm singlet} - Z_\Gamma^{\rm nonsiglet}$,
  for $\Gamma = S$ (top panels) and $\Gamma = AV$ (bottom), as a function of
  $c_{SW}$ (left) and $\omega$ (right). Parameter values relevant for
  ETMC action ($N_f = 4$, Iwasaki gluons, $\beta = 1.95$).}
\label{ETMCplots}
\end{center}
\end{figure}

\begin{figure}[h]
\begin{center}
\epsfig{figure=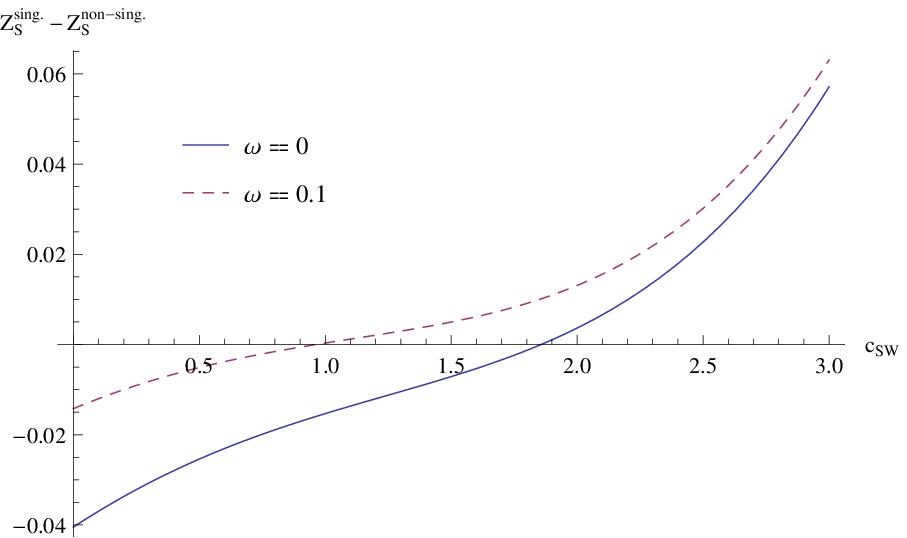,scale=0.6}
\hspace{.05\textwidth}
\epsfig{figure=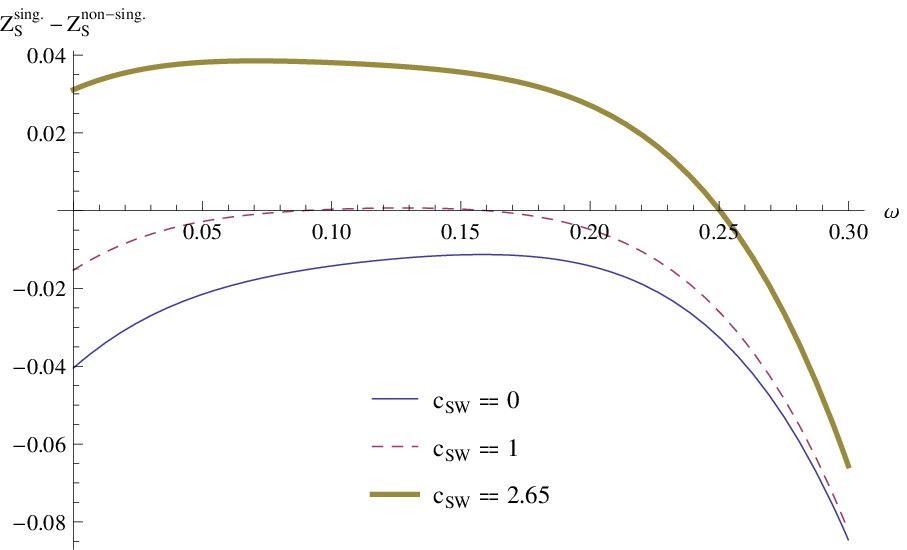,scale=0.6}\\[2mm]
\epsfig{figure=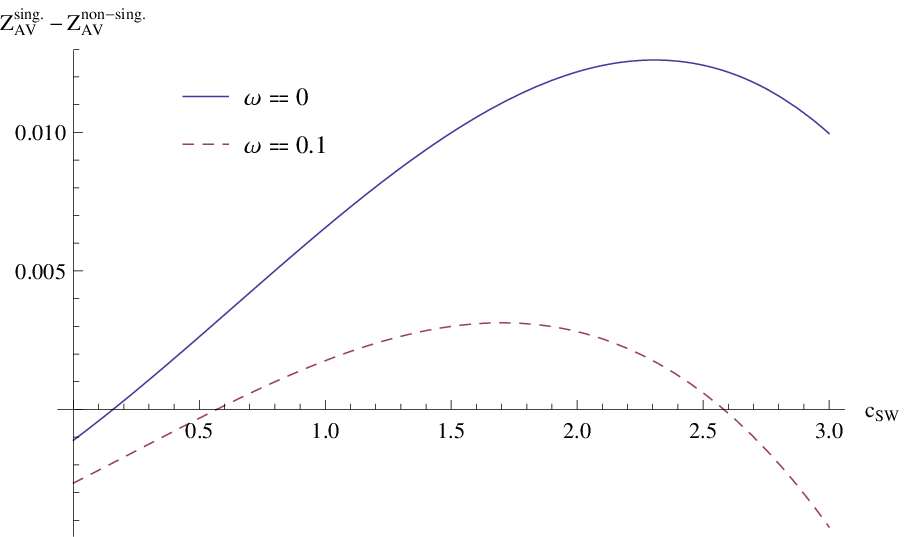,scale=0.6}
\hspace{.05\textwidth}
\epsfig{figure=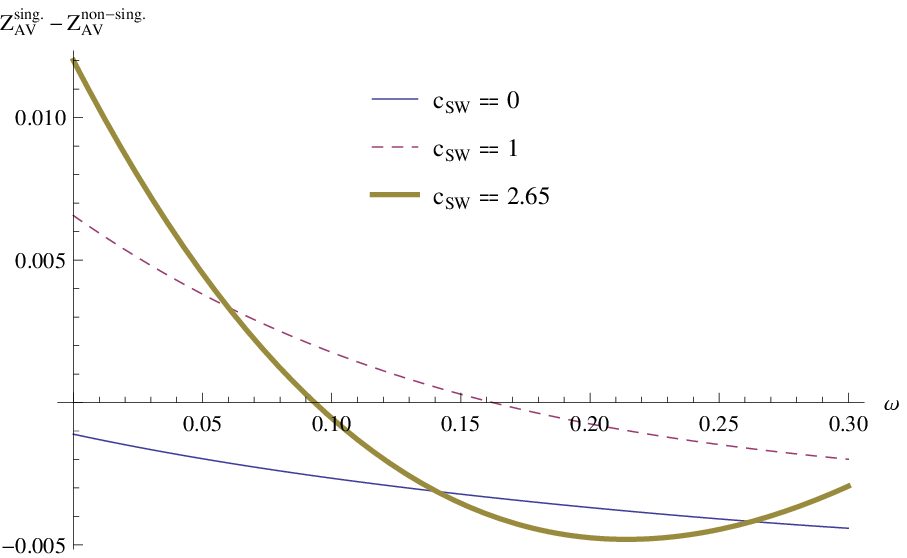,scale=0.6}

\caption{Plots of $Z_\Gamma^{\rm singlet} - Z_\Gamma^{\rm nonsiglet}$,
  for $\Gamma = S$ (top panels) and $\Gamma = AV$ (bottom), as a function of
  $c_{SW}$ (left) and $\omega$ (right). Parameter values relevant for
  SLiNC action ($N_f = 3$, Tree level Symanzik gluons, $\beta = 5.5$).}
\label{SLiNCplots}
\end{center}
\end{figure}

In Fig.~\ref{ETMCplots} we illustrate our results by selecting
parameter values appropriate to the
ETMC action with Iwasaki gluons, $N_f = 2+1+1$, $\beta = 2 N_c/g_\circ^2 = 1.95$,
$\bar\mu = 1/a_{_{\rm L}}$ and standard/stout links for the Wilson part of the
fermion action. Similarly, Fig.~\ref{SLiNCplots} presents our results for
parameter values appropriate to the SLiNC action, with Tree-Level
Symanzik gluons, $N_f = 3$,  $\beta = 2 N_c \Red{c_0}/g_\circ^2 =
5.5$, $\bar\mu = 1/a_{_{\rm L}}$\,.

Further extensions of the present work include the application to
other actions currently used in numerical simulations, including
actions with more steps of stout smearing. In these cases, 
additional contributions to the renormalization functions are more
convergent, and thus their perturbative treatment is simpler;
nevertheless, the sheer size of the vertices ($\sim 10^6$ terms for
two stout-smearing steps) renders the computation quite cumbersome.
Another possible extension of this work regards 
several variants of staggered fermion actions. Finally, extended
versions of $\bar\psi\Gamma\psi$ may be studied; in this case another
four Feynman diagrams must be added to those of Fig.~\ref{Feynman},
leading to longer but convergent loop integrands.

\bigskip\noindent
{\bf Acknowledgments:} Work supported by the Cyprus Research Promotion
Foundation under Contract 
No. TECHNOLOGY/${\rm \Theta E\Pi I\Sigma}$/0311(BE)/16.


\begin{thebibliography}{99}
\bibitem{CHP} M.~Constantinou, M.~Hadjiantonis and H.~Panagopoulos,
  \emph{Renormalization of Flavor Singlet Fermion Operators with
    Improved Actions}, in preparation.

\bibitem{FH} A.~J.~Chambers, R.~Horsley, Y.~Nakamura et al., 
\emph{A novel approach to nonperturbative renormalization of singlet
  and nonsinglet lattice operators}, {\tt arXiv:1410.3078}\,.

\bibitem{SP} A.~Skouroupathis, H.~Panagopoulos, 
\emph{Two-loop renormalization of scalar and pseudoscalar fermion
  bilinears on the lattice}, \emph{PRD} {\bf 76} (2007) 094514
[{\tt arXiv:0707.2906}];
\emph{Two-loop renormalization of vector, axial-vector and tensor
  fermion bilinears on the lattice}, \emph{PRD} {\bf 79} (2009) 094508
[{\tt arXiv:0811.4264}].

\bibitem{DiRenzo} M.~Brambilla, F.~Di~Renzo, M.~Hasegawa,
\emph{High-loop perturbative renormalization constants for Lattice
  QCD}, {\tt arXiv:1402.6581}\,.
 
\end{thebibliography}
\end{document}